# Structural investigation of the Sm(Fe$_{1-x}$Ru$_x$)As(O$_{0.85}$F$_{0.15}$) system: a synchrotron X-ray powder diffraction study

A. Martinelli[1,*]

[1] SPIN-CNR, C.so Perrone 24, 16152 Genova – Italy

**Abstract**

The structural and microstructural properties of the Sm(Fe$_{1-x}$Ru$_x$)As(O$_{0.85}$F$_{0.15}$) system were investigated by means of high-resolution synchrotron X-ray powder diffraction between 10 K and 300 K. The tetragonal to orthorhombic structural transition temperature decreases with the increase of the Ru content and the symmetry breaking is completely suppressed around $x \sim 0.38$. By combining the present results with previous magnetic and resistivity measurements, a phase diagram for the Sm(Fe$_{1-x}$Ru$_x$)As(O$_{0.85}$F$_{0.15}$) system has been drawn.

**1. Introduction**

After the discovery of superconductive properties [1], the class of materials referred to as Fe-based superconductors attracted a huge amounts of studies in the last decade. In particular compounds belonging to the *Ln*FeAsO family (*Ln*: lanthanide) exhibit the highest superconductive transition temperatures in bulk materials [2]; normally, the superconductivity is activated by electron doping in (*i.e.* partial substitution of O with F), but also hole-doping is effective. The ground state is magnetic in the undoped *Ln*FeAsO compounds, even though its nature is still debated [3,4]; in fact it is not clear if magnetism is itinerant, related to a spin density wave, or due to local magnetic moments, set within a classical antiferromagnetic spin ordering. As the pure compound is doped, the long-range magnetic order gradually turns into a short-range order; then a crossover region is observed, where the superconductivity and magnetism coexist at a nanoscopic scale before the establishment of a pure superconductive state [5].

At room temperature *Ln*FeAsO compounds crystallize in the tetragonal system, but on cooling a tetragonal to orthorhombic structural transition takes place at $T_s$, a few degrees above the magnetic transition ($T_m$) [6,7]. In the hole-doped compounds the structural transition is not affected by the degree of chemical substitution and the pure and optimally doped compounds exhibit nearly the same $T_s$ [8]. Conversely, most investigations report a progressive suppression of the structural transition with the increase of the electron doping [9,10,11,12,13]. Nonetheless, closer structural analyses carried out on SmFeAs($O_{1-x}F_x$) compounds by synchrotron X-ray powder diffraction revealed that F-substitution actually decreases the amplitude of the orthorhombic distortion (making very difficult its detection for relatively high level of F content) but not completely suppresses the structural transition [14,15]; this conclusion was then confirmed by a NMR analysis [16].

---

[*] Corresponding author: alberto.martinelli@spin.cnr.it

The origin of the structural transition is not yet clear; it is generally ascribed to orbital or spin degrees of freedom [17,18,19,20,21], since no displacive lattice degrees of freedom are involved [22]. The occurrence of a static incommensurate modulated structure developing across the low-temperature orthorhombic phase of La($Fe_{1-x}Mn_x$)AsO samples recently suggested that a charge-density-wave instability can play a primary role in determining the structural, magnetic and transport properties of Fe-based superconductors [23].

The correct definition of the phase diagrams of these systems is crucial because of the close interplay between the crystallo-chemical and magnetic properties as well as the possible coexistence of magnetism and superconductivity [24]. At this scope, chemical substitution provides one of the most effective methods to investigate the relationships among the different properties of a material. The isoelectronic Ru substitution at Fe-site can provide useful hints to highlight the correlation among these properties, since the structural, magnetic and transport properties in these materials appear to be uniquely driven by the Fe sub-lattice.

Conversely to what observed in the 122-type compounds, superconductivity cannot be achieved in $Ln$FeAsO systems by Ru-substitution [25]. In the undoped Pr($Fe_{1-x}Ru_x$)AsO system (that is a system with no electron- or hole-doping) a complete suppression of the structural transition is observed for $0.33 < x < 0.40$ [26], whereas anomalies in the transport properties which may be attributed to magnetism transition persist up to $x \sim 0.67$ [27]. Similar results were obtained for the homologous La($Fe_{1-x}Ru_x$)AsO system, where the character of the structural transition change in nature, from first to second order [28]. In addition long-range ordered magnetism occurs within the orthorhombic phase ($x \leq 0.30$), whereas short-range magnetism appears to be confined within the lattice strained region of the tetragonal phase [28].

Theoretical calculations reveal that in the $Ln$($Fe_{1-x}Ru_x$)AsO systems the chemical substitution progressively frustrates Fe moment since Ru atoms do not sustain any magnetic moment; conversely the electronic structure is only slightly affected by Ru substitution around the Fermi level [29]. Remarkably, the Sm($Fe_{1-x}Ru_x$)As($O_{0.85}F_{0.15}$) system is characterized by a re-entrant static short ranged magnetic order which degrades the superconducting ground state, due to the competition between two different order parameters, producing a nanoscopic phase separation [30]. Experimental evidences for a nanoscale electronic phase separation in this system were obtained by As K-edge extended X-ray absorption fine structure analysis [31] and high-resolution X-ray absorption and X-ray emission spectroscopy [32].

From the structural point of view, extended X-ray-absorption fine-structure measurements revealed that local disorder induced by the Ru substitution is mainly confined within the FeAs layer [33]. Interestingly, Ru displays a tendency toward local aggregation and the formation of relatively extended of Fe-enriched zones were detected by coupling pair distribution function data [34] with $^{75}$As NQR analysis [35].

The aim of the present work is to draw the phase diagram of the $Sm(Fe_{1-x}Ru_x)As(O_{0.85}F_{0.15})$ system by connecting the structural information obtained by means of high resolution X-ray powder diffraction (reported hereinafter) with magnetic and superconductive data previously obtained on the same samples [29,30,31,32,33].

## 2. Experimental

Poly-crystalline $Sm(Fe_{1-x}Ru_x)AsO(O_{0.85}F_{0.15})$ ($0.00 \leq x \leq 0.50$) samples were prepared reacting stoichiometric amounts of pre-synthesized SmAs with high purity $Fe_2O_3$, $RuO_2$, $FeF_2$, Fe, Ru [29]. Synchrotron X-ray powder diffraction (XRPD) analysis was carried out on selected samples ($x = 0.05$, 0.10, 0.20, 0.30 0.36, 0.50) at the ID31 beam-line ($\lambda = 0.35443$ Å) of the European Synchrotron Radiation Facility (ESRF) in Grenoble; for each sample XRPD patterns were collected at 300, 250, 200, 150, 100, 50 and 10 K. Thermo-diffractograms of the 110 diffraction peak (tetragonal indexing) were also acquired on heating in a continuous scanning mode (the tetragonal 110 peak splits into the 020 + 200 orthorhombic lines on cooling, marking the symmetry breaking).

Structural refinement was carried out according to the Rietveld method [36] using the program FULLPROF [37]; refinements were carried out using a file describing the instrumental resolution function. In the final cycle the following parameters were refined: the scale factor; the zero point of detector; the background (parameters of the 5$^{th}$ order polynomial function); the unit cell parameters; the atomic site coordinates not constrained by symmetry; the atomic displacement parameters; the anisotropic strain parameters. On the basis of the refined anisotropic strain parameters, the microstructural analysis was carried out and the tensor isosurfaces representing the anisotropic microstrain distribution along the different crystallographic directions were calculated.

Amounts of these same samples were previously analysed by electrical resistivity, Hall effect, magnetoresistivity measurements [29], muon spin rotation analysis [30], X-ray absorption and X-ray emission spectroscopy [32,33]. In particular, $^{19}$F nuclear magnetic resonance measurements revealed that the relative fluorine content is constant within $\Delta \leq 0:01$ in the whole set of investigated samples [30].

## 3. Results and Discussion

### 3.1 Structural refinement

At 300 K all the samples crystallize in the tetragonal $P4/nmm$ space group; no evidence for ordering at the transition metal site between Ru and Fe atoms (possibly revealed by super-lattice reflections) can be detected. The XRPD patterns reveal the presence of a few amounts of SmOF in all the examined samples. As already reported [29], the lattice parameters exhibit opposite trends when plotted as a function of the Ru content, similarly to what observed in the undoped $La(Fe_{1-x}Ru_x)AsO$ system [28]. Both lattice parameters follows rather linear behaviours, suggesting a tendency towards structural relaxation [38], in agreement with X-ray-absorption spectroscopy analyses carried out on the same samples, that measured a difference between Fe-As and Ru-As bond lengths of ~ 0.03 Å, half than expected [33].

Table 1: Structural parameters of $Sm(Fe_{1-x}Ru_x)As(O_{0.85}F_{0.15})$ samples, as refined from XRPD data collected at 300 K; space group $P4/nmm$, origin choice 2; Sm and As atoms at $2c$ site, Fe and Ru atoms at $2b$ site, O and F at atoms at $2a$ site.

|        | $x = 0.05$ | $x = 0.10$ | $x = 0.20$ | $x = 0.30$ | $x = 0.36$ | $x = 0.50$ |
|--------|-----------|-----------|-----------|-----------|-----------|-----------|
| $a$ (Å) | 3.9383(1) | 3.9439(1) | 3.9544(1) | 3.9638(1) | 3.9735(1) | 3.9956(1) |
| $c$ (Å) | 8.4872(1) | 8.4872(1) | 8.4571(1) | 8.4348(1) | 8.4018(1) | 8.3429(1) |
| $z$ Sm  | 0.1410(1) | 0.1392(1) | 0.1389(1) | 0.1384(1) | 0.1388(1) | 0.1385(1) |
| $z$ As  | 0.6608(1) | 0.6606(1) | 0.6604(1) | 0.6607(2) | 0.6606(1) | 0.6610(1) |
| $\chi^2$ | 3.17      | 4.16      | 4.24      | 6.81      | 5.95      | 3.51      |

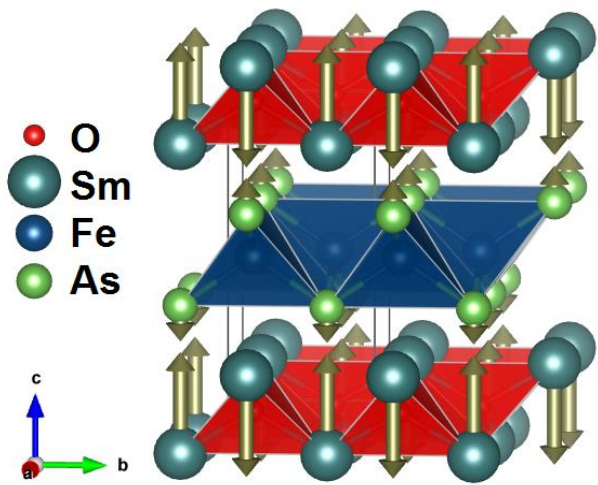

Figure 1: Distortion of the $SmFeAs(O_{0.85}F_{0.15})$ crystal structure induced by Ru-substitution.

Figure 1 shows the effect of Ru substitution on the crystal structure of $SmFeAs(O_{0.85}F_{0.15})$: the slight expansion along the $c$-axis in the tetrahedral Fe-As layer induced by Ru-substitution is coupled with a notable compression of the Sm-O layer. This behaviour can be explained as follows: the Ru ions being larger than Fe ions produce the observed expansion of the Fa-As layer along the *c*-axis. At the same time Ru is a 5*d* Ru ion and hence is its electrons are more delocalized the those of the 3*d* Fe ion. In pure SmFeAsO, As atoms are negatively charged and receive negative charge from both Fe and Sm atoms [39]. When the 3*d* Fe ion is substituted by the 5*d* Ru ion, negative charge can is thus more easily transferred from the Fe plane to the As plane. As a consequence, the electrons of the Sm atoms are more effectively attracted by the plane constituted of O atoms and hence the Sm-O layer compresses.

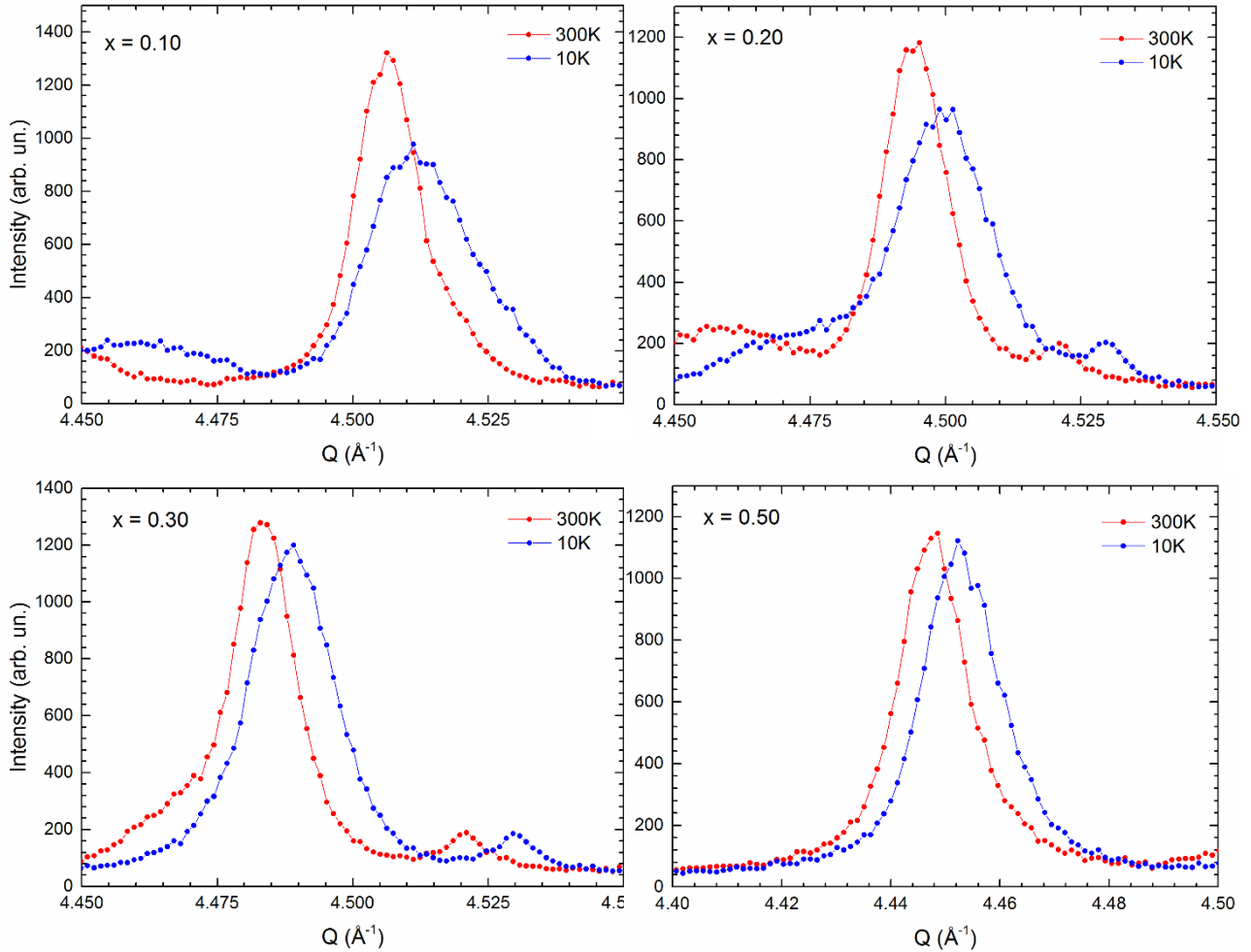

Figure 2: Selected regions of the XRPD patterns (samples with $x$ = 0.10, 0.20, 0.30) showing the evolution on temperature of the tetragonal 220 diffraction line (data at 300 K) into a broadened peak at low temperature (10 K) constituted by the convolution of the orthorhombic 400 and 040 diffraction lines; for the sample with $x$ = 0.50 only a very small broadening can be detected.

The pure compound SmFeAsO undergoes a *P*4/*nmm* → *Cmme* structural transition on cooling that is evidenced by the splitting of the tetragonal *hh*0 diffraction lines into orthorhombic *h*00+0*k*0 lines [7,40]. Nonetheless, F-substitution progressively decreases the degree of the orthorhombic distortion [14,40] and Ru-substitution also provides a similar effect, as already observed in the homologous La(Fe$_{1-x}$Ru$_x$)AsO system [28]; moreover, chemical substitution intrinsically broadens the diffraction lines. For these reasons, the decreased orthorhombic distortion and the broadening of the diffraction peaks are concurrent issues that hinder the development of a clear peak split; rather, these issues give rise to a single very broadened peak resulting from the convolution of the unresolved orthorhombic diffraction lines. Figure 2 shows selected regions of the XRPD patterns for different samples. At 300 K the full width at half maximum (FWHM) of the tetragonal 200 peak is comparable for all the inspected samples, ranging between ~0.013 and ~0.014 Q for $x \leq 0.36$ (the instrumental contribution to peak broadening in this region of the pattern is ~0.001 Q). At 10 K the FWHM value for $x = 0.10$ is almost doubled, being 0.022 Q. Such a huge broadening could suggest that the peak at low temperature is the convolution of the orthorhombic 400+040 diffraction lines, as confirmed by Rietveld refinement (*vide infra*). Qualitatively similar results are obtained for samples with $x \leq 0.36$, although the degree of broadening progressively decreases with the increase of the Ru content, as expected because of the reduction of the orthorhombic distortion. Also the sample with $x = 0.50$ displays a faint broadening at 10 K, but its origin can be ascribed to distortions confined to a local scale that do not produce a structural transition; in fact Rietveld refinements indicate that in this sample the tetragonal structural model better fits the experimental data in the whole inspected thermal range.

Structural refinements using data collected at low temperature were carried out applying both tetragonal and orthorhombic structural models, taking into account also the anisotropic strain broadening contribution. By comparing the corresponding weighted $\chi^2$ values (listed in Table 2 for data at 10 K), it is found that the orthorhombic model fits the data better than the tetragonal one in several cases ($\chi^2 \geq 1$ because of the very high precision synchrotron XRPD data [41]).

In order to ascertain if the $\chi^2$ decrease is related to a real improvement of the structural model, the significance test on the crystallographic $R$ factor was applied [42]; in particular, this test statistically assesses the improvement of the fit when the structural model is changed. The number of diffraction peaks in the inspected 2θ range are 406 for the tetragonal structural model, but those characterized by a real detectable intensity are only 221 (relative intensity more than about 1.5). In the last refinement cycle, the orthorhombic structural model has 3 parameters more than the tetragonal one (1 cell parameter plus 2 anisotropic strain parameters) and hence the dimension of our hypothesis is 3. The number of degrees of freedom for the refinement is 197, that is the difference between the number of the diffraction lines (221) and the number of the refined parameters in the orthorhombic structural model (24). The value $\mathscr{R}$ given by the ratio between the tetragonal and orthorhombic $\chi^2$ is compared with the value of significance points for $\mathscr{R}$ (for a significance level α = 0.005: $\mathscr{R}_{3,197,0.005}$), that can be obtained by interpolation [43], resulting $\mathscr{R}_{3,197,0.005} = 1.0338$. In other words, the tetragonal structural model can be rejected at the 0.5% level of significance when $\mathcal{R}$ is grater of 1.0338. This analysis indicates that the structural transition is still active in the samples with $x \leq 0.36$, with the transition temperature decreasing with the increase of the Ru content.

Noteworthy, the improvement gained for the orthorhombic structural model actually results from a better fitting of the diffraction lines with a strong component in the tetragonal *xy* plane (tetragonal *hhl* diffraction lines). This is quite evident in Figure 3 showing the comparison of the XRPD data collected at 10 K on the $x = 0.05$ sample fitted with the *P*4/*nmm* and *Cmme* structural models; the tetragonal model overestimates the intensities of the peak at ~ 2.26 Å$^{-1}$, whereas the orthorhombic model perfectly fits the diffraction line.

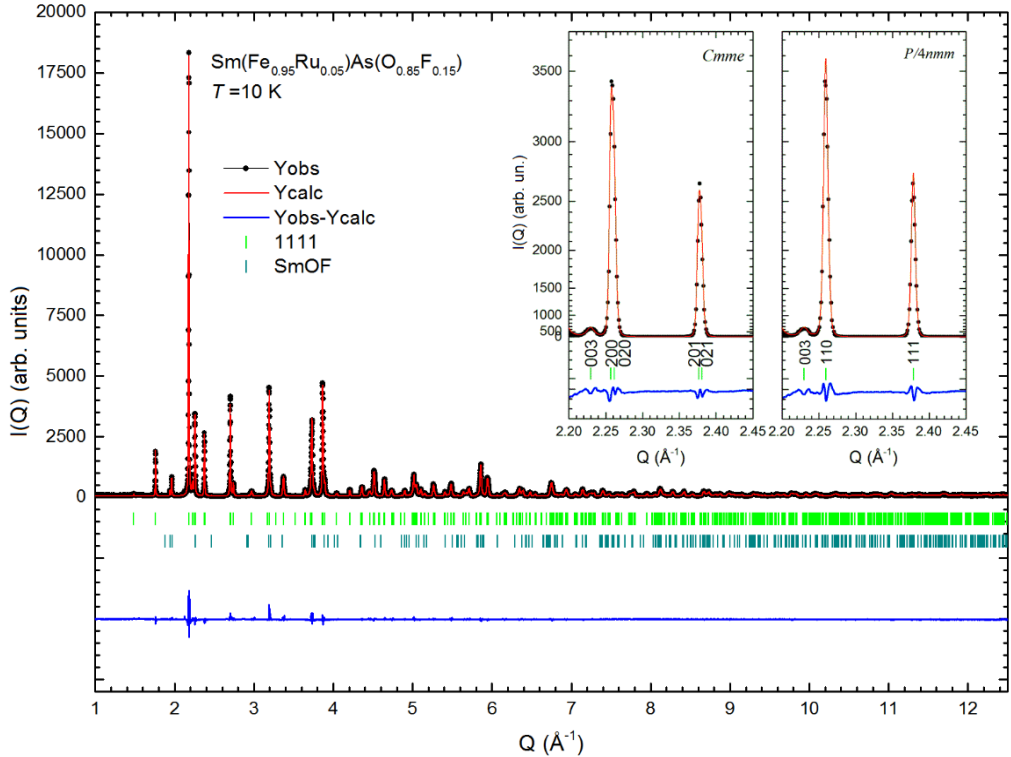

Figure 3: Rietveld refinement plot obtained fitting the XRPD data collected at 10 K on the Sm(Fe$_{0.95}$Ru$_{0.05}$)As(O$_{0.85}$F$_{0.15}$) sample with an orthorhombic structural model; tick marks indicate the position of the Bragg peaks (including those of SmOF), the points are observed data, whereas the solid line is the calculated profile; a difference curve (observed minus calculated) is plotted at the bottom. The inset shows portions of full-pattern Rietveld fits obtained with a tetragonal and orthorhombic structural models in the **Q** range 2.20-2.45 Å.

The significant changes affecting the tetragonal *hhl* diffraction lines at low temperature are also observed in other samples with increased Ru-content. Figure 4 (on the left) shows the thermal evolution of the coherent X-ray scattering in the region where the tetragonal 110 diffraction line splits into the 020 + 200 orthorhombic lines on cooling. It is evident the progressive decrease of the peak intensity on cooling, coupled with a line broadening increase; these features point to a structural change. Figure 4 (on the right) shows the evolution with temperature of the integral breadth and the observed intensity (raw data). The integral breadth data collected in the 175-300 K thermal range can be linearly fitted, but data gradually deviates from this behaviour at lower temperatures. At first, the peak broadens due to the increase of the lattice strains as the structural transition is approached [15]; subsequently, the deviation markedly increases after the symmetry breaking, because the peak is actually a convolution of two different diffraction lines that progressively split up on cooling. At the same time, the observed intensity quite linearly decreases on cooling down to ~ 125 K; then it remains quite constant in the thermal range 90 – 125 K and definitely reduces as the temperature is further decreased, a behaviour that strongly points to a significant structural reorganization.

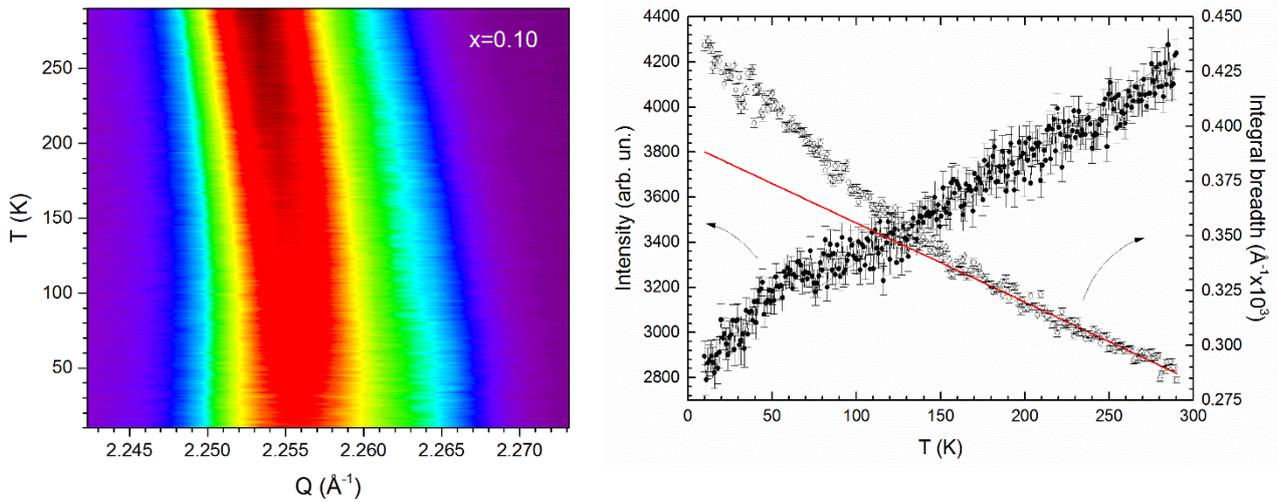

Figure 4: Sample Sm(Fe$_{0.90}$Ru$_{0.10}$)As(O$_{0.85}$F$_{0.15}$). On the left: Thermo-diffractograms showing the thermal evolution of the coherent X-ray scattering in the region of the tetragonal 110 diffraction line. On the right: thermal evolution of the corresponding integral breadth (the continuous line is the linear fit of the data collected between 175 K and 300 K) and the observed intensity.

Figure 5 shows the evolution of the cell parameters for selected compositions, whereas in Table 2 are reported the structural data at 10 K for all the analysed samples.

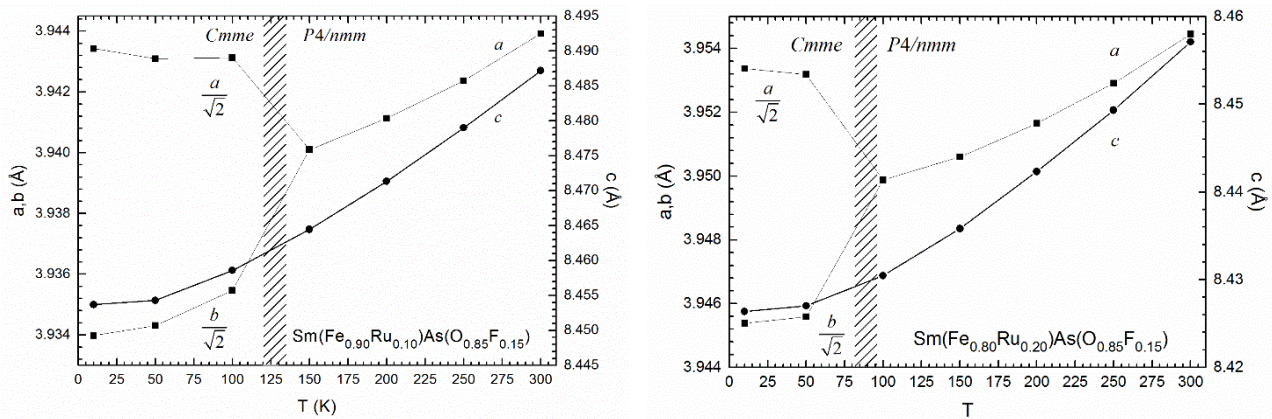

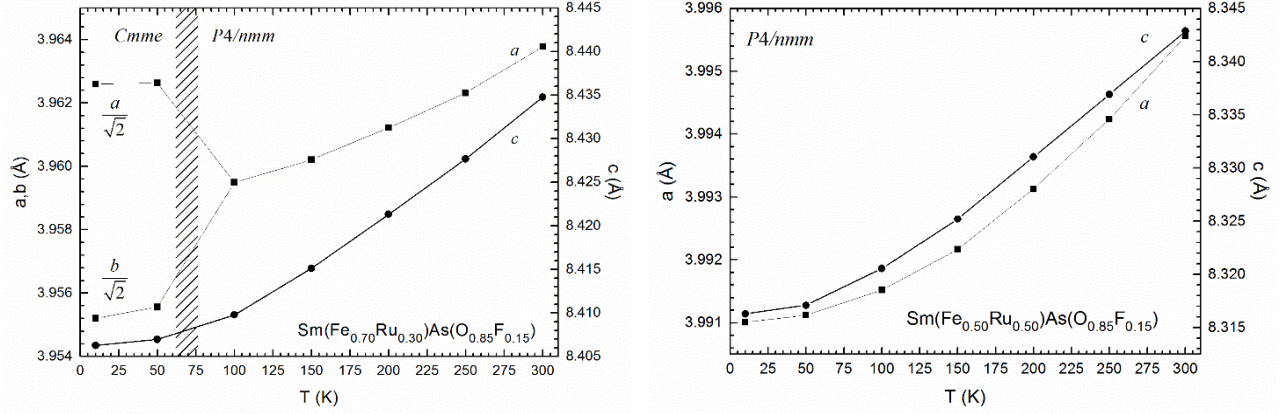

Figure 5: Thermal dependence of the cell parameters in some of the analysed samples (lines are guide to eye); the hatched region represents the thermal range where the symmetry breaking takes place.

Table 2: Structural parameters of $Sm(Fe_{1-x}Ru_x)As(O_{0.85}F_{0.15})$ samples, as refined from XRPD data collected at 10 K. Samples with $0.10 \leq x \leq 0.36$ crystallize in the *Cmme* space group; Sm and As atoms at 4$g$ site, Fe and Ru atoms at 4$b$ site, O and F at atoms at 4$a$ site. The sample with $x = 0.50$ crystallizes in the *P4/nmm* space group).

|  | $x = 0.05$ | $x = 0.10$ | $x = 0.20$ | $x = 0.30$ | $x = 0.36$ | $x = 0.50$ |
|---|---|---|---|---|---|---|
| $a$ (Å) | 5.5677(1) | 5.5768(1) | 5.5909(1) | 5.6040(1) | 5.6172(1) | 3.9910(1) |
| $b$ (Å) | 5.5563(1) | 5.5635(1) | 5.5796(1) | 5.5935(1) | 5.6080(1) | / |
| $c$ (Å) | 8.4543(1) | 8.4537(1) | 8.4264(1) | 8.4063(1) | 8.3739(1) | 8.3163(1) |
| $z$ La | 0.1410(1) | 0.1392(1) | 0.1392(1) | 0.1386(1) | 0.1390(1) | 0.1387(1) |
| $z$ As | 0.6606(1) | 0.6606(1) | 0.6603(1) | 0.6604(1) | 0.6605(1) | 0.6608(1) |
| $\chi^2_{orthorhombic}$ | 4.75 | 5.87 | 5.59 | 5.76 | 6.08 | 4.76 |
| $\chi^2_{tetragonal}$ | 5.13 | 6.49 | 5.90 | 6.00 | 6.29 | 4.86 |

The volume of the primitive cell decreases homogeneously on cooling, similarly to what observed in the undoped La(Fe,Ru)AsO system [28]. The thermal expansion behaviour of the samples has been investigated by fitting the cell volume between 10 and 300 K using a Grüneisen second-order approximation for the zero-pressure equation of state [44]:

$$V(T) = \frac{V_0 U}{Q - bU} + V_0 \quad (1)$$

Where $Q = V_0 K_0 / \gamma'$ and $b = (K_0' - 1)/2$; $\gamma'$ is a dimensionless Grüneisen parameter of the order of unity; $K_0$ is the compressibility and $K_0'$ its derivative with respect to applied pressure; $V_0$ is the zero temperature limit of the unit cell volume; $U$ is the internal energy calculated by the Debye approximation:

$$U(T) = 9Nk_B T \left(\frac{T}{\Theta_D}\right)^3 \int_0^{\frac{T}{\Theta_D}} \frac{x^3 dx}{e^x - 1} \quad (2)$$

Where N is the number of atoms in the unit cell; $k_B$ is the Boltzmann's constant; $\Theta_D$ is the Debye temperature. The fitting was carried out assuming $K_0 = 1.03 \cdot 10^{11}$ Pa, which is the experimental bulk modulus value extracted from high pressure XRPD measurements on SmFeAs($O_{0.93}F_{0.07}$) [45], and leaving $\gamma'$, $\Theta_D$ and $K'_0$ as free parameters. Figure 6 shows the resulting fitting curves for the inspected samples, evidencing that the Grüneisen law correctly predicts their thermal expansion behaviour. For all samples, $\Theta_D$ values range between 260 K and 315 K, in very good agreement with the values similarly obtained in the homologous series La($Fe_{1-x}Ru_x$)AsO [28] and that calculated in pure LaFeAsO ($\Theta_D = 282$ K) from heat capacity data [46].

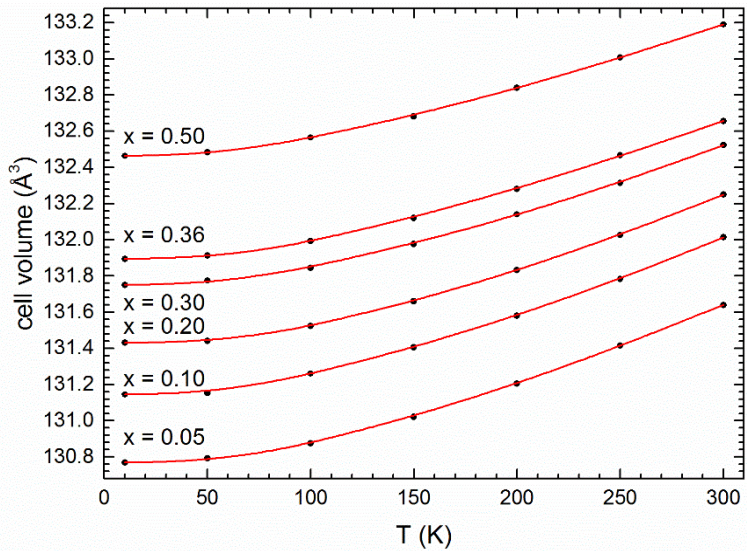

Figure 6: Unit cell volume as a function of temperature for the examined samples; the solid lines show the best fit to a second-order Grüneisen approximation.

*3.2 Microstructural analysis*

The micro-structure of the samples was investigated by using the anisotropic strain parameters obtained after Rietveld refinement and analyzing the broadening of diffraction lines by means of the Williamson-Hall plot method [47]. In general, in the case where size effects are negligible and the micro-strain is isotropic, a straight line passing through all the points in the plot and the origin has to be observed, where the slope provides the micro-strain: the higher the slope the higher the micro-strain. If the broadening is not isotropic, size and strain effects along particular crystallographic directions can be obtained by considering different orders of the same reflection.

Figure 7 shows the evolution of the micro-structural strain along the three main crystallographic directions $h00$, $hh0$ and $00l$ as obtained applying the tetragonal structural model during the Rietveld refinement in the whole inspected temperature range; for a better comparison the values reported in the figure are normalized to those calculated at 300 K. It is evident that the samples with $x \leq 0.30$ display a similar behaviour: lattice micro-strains along $h00$ and $00l$ are almost coincident in the whole thermal range, whereas micro-strain along $hh0$ departs from them on cooling. For example in the sample with $x = 0.10$ a detectable departure is observed already at 200 K, related to the taking place of lattice micro-strain within the tetragonal structure. The departure increases as the temperature is further lowered, where the orthorhombic polymorph becomes stable. In this thermal range the instrumental resolution is not sufficient to describe the splitting of the orthorhombic $h00$ and $0k0$ diffraction lines (Figure 3, inset), on account of the very reduced orthorhombic distortion. As a consequence, in the tetragonal structural model these diffraction lines are convoluted in a single $hh0$ peak, whose broadening is modelled as lattice microstrain that progressively increases with the decrease of temperature. Conversely, micro-strain along $h00$ and $00l$ tends to level off. It is interesting to observe that the samples with $x \leq 0.30$ exhibit a similar increase of the micro-strain along $h00$ and $00l$ on cooling, up to ~ 1.20 -1.25; conversely, the increase of the micro-strain along $hh0$ measured at 10 K decreases with the increase of the Ru content, as expected for a progressive reduction of the orthorhombic distortion. The thermal dependence of the microstructural properties of these samples are similar to that observed in pure SmFeAsO [15].

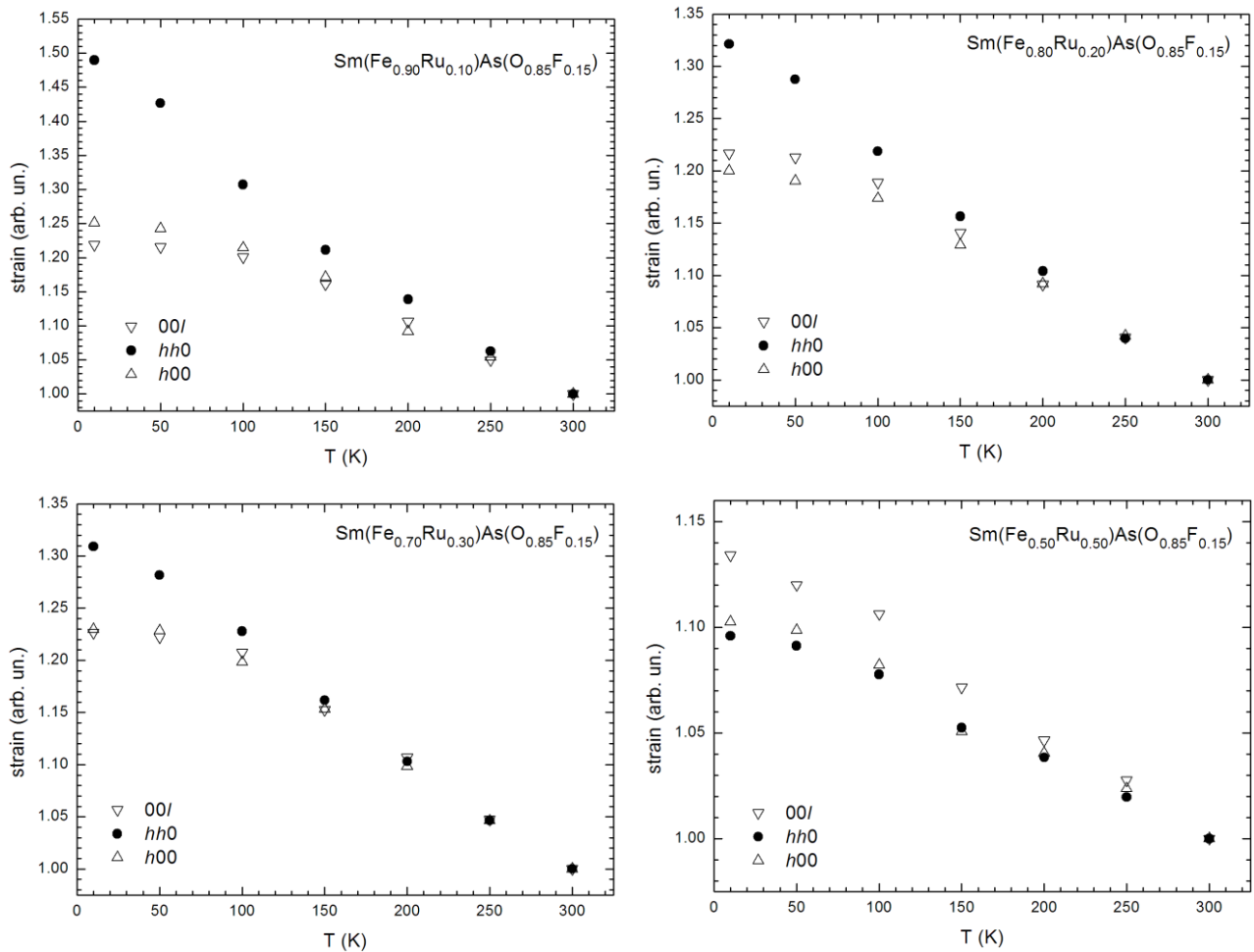

Figure 7: Thermal evolution of the micro-structural strain (normalized to the value at 300 K) along three main crystallographic directions in samples with $x = 0.10, 0.20, 0.30$ and $0.50$; refinements were carried out using a tetragonal structural model in the whole temperature range.

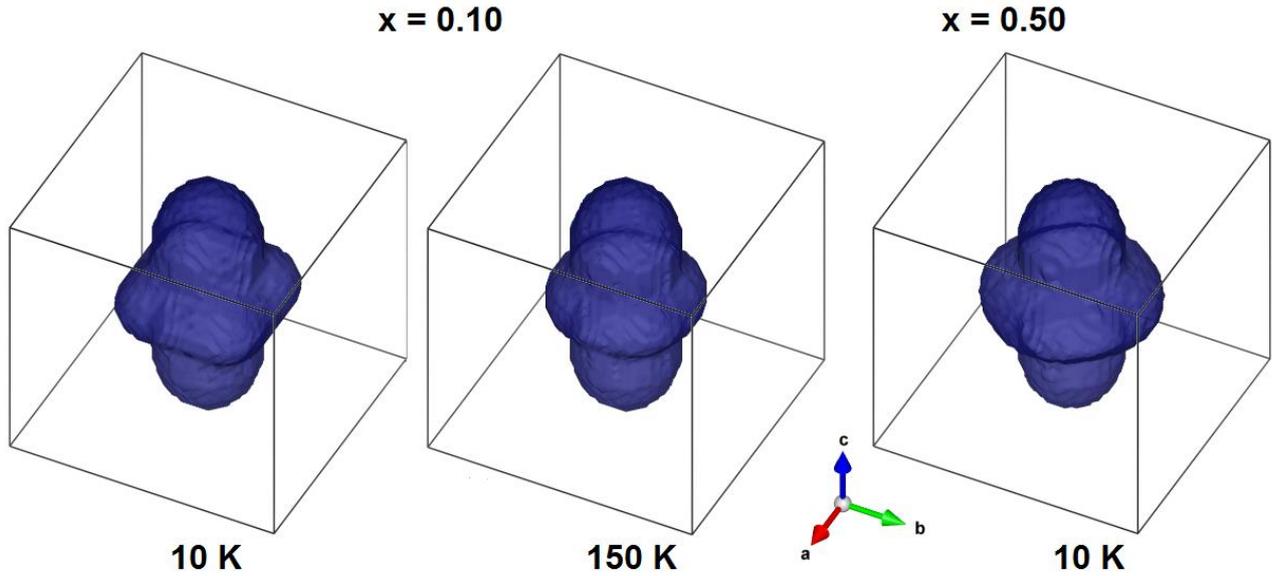

Figure 8: Experimentally observed tensor isosurfaces representing the microstrain broadening characterizing samples with $x = 0.10$ and 0.50 at low temperature (obtained applying the tetragonal structural model).

As a matter of fact, the selective diffraction line broadening (that is modelled as lattice microstrain in the tetragonal structural model) could actually mask a weak orthorhombic lattice distortion. In order to gain insights, it is necessary to ascertain if the geometrical features of the tensor isosurfaces representing the tetragonal microstrain are consistent with those expected for a $P4/nmm \rightarrow Cmme$ structural transition. Figure 8 displays the microstrain broadening observed in samples with $x = 0.10$ and 0.50 as obtained by applying the tetragonal structural model. Both samples display large strain along the $c$ axis, which is related to the distortion of the crystal structure produced by Ru-substitution, as discussed in §3.1. At 150 K, the sample with $x = 0.10$ shows an isotropic distribution of the microstrain in the $ab$ plane; at 10 K an in-plane 4-fold tensor surface is instead observed, consistent with the microstrain distribution expected for a $4/mmm \rightarrow mmm$ structural transition [48]. This same anisotropy is observed also in other systems where a tetragonal-to-orthorhombic structural transformation (involving a point group $4/mmm \rightarrow mmm$ transition) takes place, such as $Pb_3O_4$ [48,49]. Conversely, the anisotropy of the in-plane microstrain broadening is negligible for $x = 0.50$ down to 10 K, indicating that in this case no structural transition takes place, in agreement with the Rietveld refinement results. In this sample the short-range fluctuations of the lattice parameters may determine a widespread orthorhombic distortion of the local crystallographic structure, but the average structure remains tetragonal in the whole thermal range.

*3.3 The phase diagram*

In Figure 9 the phase diagram for the $Sm(Fe_{1-x}Ru_x)As(O_{0.85}F_{0.15})$ system is drawn, based on the experimental results of previous investigations [29,30,31,32,33] and the present structural analysis.

Unfortunately, from our data it is not possible to ascertain the order of the structural transition and its possible change as a function of the Ru-content.

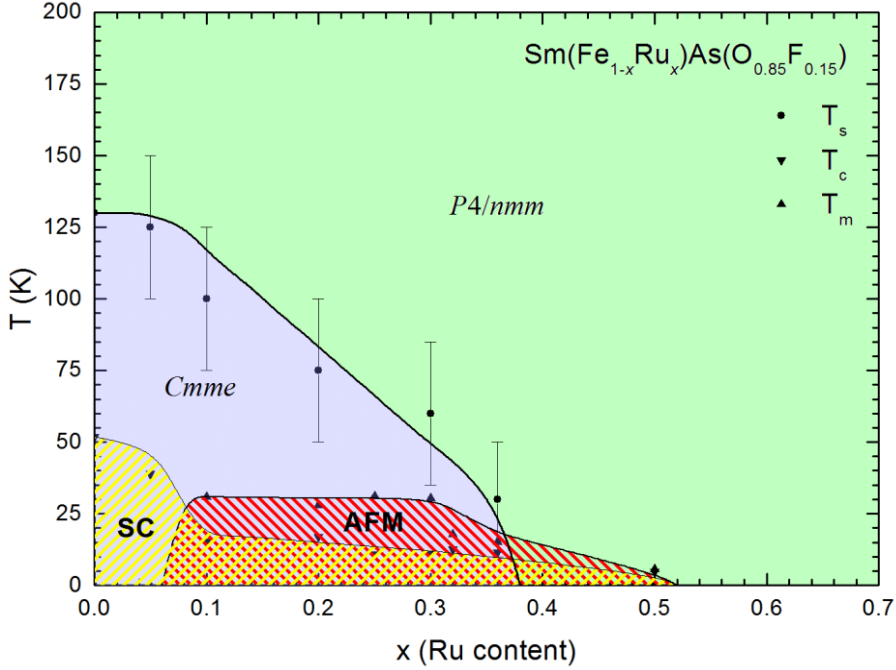

Figure 9: Phase diagram of the Sm(Fe$_{1-x}$Ru$_x$)As(O$_{0.85}$F$_{0.15}$) system; the solid line separates the tetragonal field form the orthorhombic one, patterned areas define the superconducting and magnetic regions.

For $x \leq 0.05$ the structural transition temperature $T_s$, is similar with that of SmFeAs(O$_{0.85}$F$_{0.15}$) [14], whereas at higher level of Ru substitution $T_s$ definitely falls down. This behaviour is likely determined by short-range chemical correlations taking place at the transition metal sub-lattice, as observed in the homologous La(Fe$_{1-x}$Ru$_x$)AsO system [34]. It is then reasonable to assume that for a low degree of substitution such correlations are negligible, since the substituent Ru ions are very diluted in the transition metal sub-lattice; the effectiveness of these correlations grows with the increase of Ru content, thus producing a sudden decrease of the structural transition temperature.

In the compositional range $0.10 \leq x \leq 0.30$ the magnetic transition temperature $T_m$ remains constant, whereas the structural transition temperature $T_s$ undergoes a remarkable decrease. This behaviour suggests that the structural and the magnetic degrees of freedom are not correlated and hence that the structural transition is not activated by magnetism. Remarkably, the values of both $T_m$ and $T_c$ remain almost constant in this same compositional range and then similarly decrease down to their suppression. This is consistent with a scenario where magnetism and superconductivity are driven by the same kind of interactions and thus compete for the same electrons, but, at the same time, coexist at the microscopic scale (unconventional $s^{+-}$ pairing state [50]).

As the Ru content further increases, the structural transition is suppressed and the magnetic ordering as well; nonetheless, magnetism endures in the tetragonal phase field, likely confined within structurally strained regions (structural distortions confined to a local scale), as in the homologous La(Fe$_{1-x}$Ru$_x$)AsO system [28]. Remarkably, the structural transition is suppressed in the La(Fe$_{1-}$ $_x$Ru$_x$)AsO [28], Pr(Fe$_{1-x}$Ru$_x$)AsO [26] and Sm(Fe$_{1-x}$Ru$_x$)As(O$_{0.85}$F$_{0.15}$) systems at about the same Ru content, suggesting that the same mechanism is at play in these systems, regardless of the different chemical pressure induced by the rare earth.

## 4. Conclusions

In conclusion, the phase diagram of the Sm(Fe$_{1-x}$Ru$_x$)As(O$_{0.85}$F$_{0.15}$) system (Figure 9) provides evidence of the fundamental role of the electronic degrees of freedom (orbital or charge order) in the activation of the structural transition, in agreement with early theoretical works [51,52]. In fact, it is evident that the establishments of the magnetic order and the orthorhombic symmetry are characterized by completely different behaviours, thus indicating that magnetic degree of freedom can not account for the structural transition. Moreover, the intimate coexistence and relationship between magnetism and superconductivity at the microscopic scale in the compositional range $0.10 \leq x \leq 0.30$ indicate that these states compete for the same electrons. Both states are suppressed as the Ru-content exceeds a critical threshold, marking their close relationship and possibly suggesting that superconductivity might be mediated by spin fluctuations.

**Acknowledgments**

A.M. acknowledges Prof. A. Palenzona for his invaluable assistance during sample preparation and Dr. Caroline Curfs for her kind support during data collection at the ID31 beamline of ESRF (proposal HS-4578).